\documentstyle[12pt]{article}
\newcommand{\half}{{1\over2}}
\newcommand{\be}{\begin{equation}}
\newcommand{\ee}{\end{equation}}
\newcommand{\bea}{\begin{eqnarray}}
\newcommand{\eea}{\end{eqnarray}}

\newcommand{\x}{\prime }

\newcommand{\zz}{L_{z}}

\newcommand{\oox}{x_{o}}
\newcommand{\iix}{x_{i}}
\newcommand{\LLx}{L}

\begin{document}

\begin{center}
{\bf One-Loop Corrected Thermodynamics of the Extremal and Non-Extremal\\}
{\bf  Spinning BTZ Black Hole\\}  
\end{center}
\begin{center}
{\sl by\\
}
\vspace*{0.40cm}
{\bf A.J.M. Medved${}^{\flat\star} $   and G. Kunstatter$^\sharp$\\}
\vspace*{0.50cm}
{\sl
$\flat$ Permanent Address:\\
 Dept. of Physics and Winnipeg Institute of
Theoretical Physics\\
 University of Manitoba, Winnipeg, Manitoba\\
Canada R3T 2N2\\
{[e-mail: joey@theory.uwinnipeg.ca]}}\\
\vspace*{0.50cm}
{\sl
$\star$
Current Address:\\
Dept. of Mathematics and Applied Mathematics\\
University of Cape Town, Rondebosch 7701\\
Cape Town, South Africa\\
{[e-mail: joey@cosmology.mth.uct.ac.za]}}\\
\vspace*{0.50cm}
{\sl
$\sharp$ Dept. of Physics and Winnipeg Institute of
Theoretical Physics\\
 University of Winnipeg, Winnipeg, Manitoba\\
Canada R3B 2E9\\
{[e-mail: gabor@theory.uwinnipeg.ca]}
}
\end{center}
\bigskip\noindent
{\large
ABSTRACT
}
\par
\noindent
We consider the  one-loop corrected 
geometry and thermodynamics of a rotating BTZ black hole
by way of a dimensionally reduced dilaton model. The analysis begins 
 with a 
comprehensive study of the non-extremal solution   after which two
different methods are invoked to study the extremal case. The first
approach considers the  extremal limit of the   non-extremal calculations,
whereas the second treatment  is based on the following  conjecture:  
extremal
and non-extremal black holes are  qualitatively distinct entities.
We show that only the latter method yields regularity and consistency 
at the one-loop level. This is suggestive of a generalized
third law of thermodynamics that forbids   continuous
evolution from non-extremal to extremal black hole geometries.
 \\
\newpage

\section{Introduction}\medskip
\par
Nearly thirty years have passed since Bekenstein and Hawking conjectured
the laws of black hole mechanics to be in analogy with those  of 
thermodynamics \cite{1}. This analogy is now  widely accepted as
an actual physical relation  rather than just a mathematical anomaly.
This   in large part is  due to Hawking's landmark discovery  that
black holes radiate thermally \cite{3}. One of the more important
open issues in this regard is the microscopic source of black hole
entropy~\cite{4}.
\par
In the case of non-extremal black holes, the quantifying relations  for the 
temperature  and entropy  are well established: $T=\kappa/2\pi$
and $S=A/4$ (respectively),  where $\kappa$ denotes the surface gravity, 
$A$ denotes the
surface area of the outer
horizon  and all fundamental constants have been set to unity. However,
when  extremal black holes  are considered (i.e., charged or spinning black
holes with a degenerate horizon singularity),  it is an entirely 
different matter. There is still no consensus  regarding  
extremal thermodynamics.
\par
 Extremal black holes  are   often  interpreted as a  limiting case of  
 non-extremal 
solutions \cite{4.50}, and this viewpoint leads to $T_{ext}=0$ and 
$S_{ext}=A_{ext}/4>0$.
However, Hawking et al.  \cite{5} and others [6-8]  have 
argued strongly against this intuitive
 notion. The ``Hawking conjecture''  that extremal and non-extremal black
holes are qualitatively distinct objects  has  
profound influences
on thermodynamics. For instance, it has been argued that
 $T_{ext}$ is  an arbitrary 
quantity. Quantitatively, this can be 
 explained by the double zero in the extremal
metric at the horizon, 
which translates to no conical singularity    
in
 the Euclidean
(i.e., imaginary time)
sector.
  Hence, there is no method for fixing the  Euclidean time 
periodicity (which is 
equivalent to the  inverse black hole temperature), contrary to
the non-extremal case \cite{6}. Qualitatively, the arbitrary temperature
can be interpreted  as a consequence of the third law of thermodynamics (i.e.,
no  system with a finite temperature  can ever reach $T=0$), which prevents
non-extremal black holes from evolving into 
 extremal ones and {\it vice versa}.
An extremal black hole  must be free to radiate at any 
 temperature so as to
 retain   its
extremal nature,  regardless of incoming radiation.
\par
Hawking has  further  proposed   that the  arbitrary temperature implies a
 vanishing
entropy. This argument is based on  qualitative differences (in the Euclidean
sector) 
  between the   extremal and non-extremal  
topologies. 
The Euclidean topology of an extremal black hole  is relatively trivial,  
and this effectively  eliminates
the usual horizon contribution (which accounts for the non-extremal entropy
\cite{6}) to the Euclidean action.
Note that
  the findings of various  other works have since supported the Hawking
conjecture [10-15].
\par
In spite of the compelling nature of the above arguments, there is still
significant opposition to this point of view. Trivedi \cite{13} and
 Loranz et al.  \cite{14} have argued that stress tensor
regularity on the horizon (in the free-falling observer frame) will be
violated unless $T_{ext}=0$. Meanwhile, the strongest case against
$S_{ext}=0$ has come from the calculations of Strominger and others \cite{15}.
They considered certain classes of  weakly coupled string theory 
(for which massive string states can be represented by extremal black holes)
and used a statistical procedure to generate $S_{ext}=A_{ext}/4$,
precisely. The same result has been obtained elsewhere  with arguments that
favor  a well-defined extremal limit. These include Ghosh and  Mitra
\cite{17}, and Kiefer  and  Louko \cite{8} 
(quantizing the system  before extremizing), as well as
 Zaslavskii \cite{18} (confining the black hole to a finite cavity  before
 extremizing).  In  still  another viewpoint,   
Wang et al. \cite{19} have proposed that
distinct extremal solutions (``Hawking's'' and ``Zaslavskii's'') can
coexist in nature.
\par
In this paper, we hope to further understanding into the
thermodynamics of extremal black holes.
 The vehicle for our investigations is  a special dilaton
model of gravity that  describes the  1+1-dimensional projection of a
rotating BTZ black hole. The BTZ black hole refers to solutions of
2+1-dimensional anti-de Sitter gravity that were  first documented by
Banados, Teitelboim and Zanelli \cite{20}. The reduction process to a 
dilaton model is based on the work of Achucarro and Ortiz \cite{21}.
The BTZ model has sparked recent interest  due to its profound connections
with
 string theory. That is, many of the black holes pertaining to string theory
have near-horizon geometries that can be 
 expressed  as BTZ $\times$ simple manifold
\cite{22}.
\par
Recently, it has been pointed out that the procedures of dimensional reduction
and quantization do not necessarily  commute \cite{NO1}. Moreover, there is a 
so-called ``dimensional-reduction anomaly'' \cite{FZ} which implies that
renormalized quantities in the unreduced theory cannot be simply
obtained by renormalizing and summing their dimensionally reduced
counterparts. Our interest in the present paper, however, is not
necessarily to derive quantitatively accurate corrections to the BTZ
black hole thermodynamics. It is to understand qualitatively the
difference between extremal and non-extremal geometries. In this
regard, the quantization of the dimensionally reduced theory may be
adequate. In any case, in the context of 1+1-dimensional  dilaton gravity, 
the dimensional-reduction anomaly can be considered as inconsequential.
\par
The remainder of the paper  proceeds as follows. In Section 2, we consider
the  non-extremal geometry, including the calculation of
 back-reaction effects 
to the  first perturbative order.
Section 3 continues the non-extremal study  with the evaluation of 
one-loop  thermodynamics  by way of a
 Euclidean action approach  \cite{6,25,32}. In
Section 4, we  investigate the extremal 
limit by applying a  limiting procedure  to the results of
 the prior two sections. 
 Section 5
considers   an alternative method for  extremal calculations that  
 reflects 
the topological differences between the  extremal and non-extremal geometries. 
This method  is 
 similar to  the approach  taken by   Buric and Radovanovic
\cite{12.99} 
in the context of Reissner-Nordstrom black holes.  Section 6 contains a
 summary of
our  findings  along with  a brief discussion.
\par
Note  that all calculations are with respect to the Hartle-Hawking vacuum
state \cite{24}. This state  can be regarded as describing 
an eternal black hole
in thermal equilibrium or (effectively) a black hole  within a thermally
reflective ``box''. 
\par
 Although the introduction has  emphasized
extremal black holes, the non-extremal results have merit on their own.
With this in mind, we make note of other  studies \cite{25.50}
 that have  considered
the  one-loop corrected   thermodynamics
of the BTZ black hole. 

\section{Non-Extremal Geometry}
Before proceeding on with the formal discussion, 
we note that  the analytical techniques of this paper
are based on a previous one-loop  study of generic dilaton gravity
\cite{26}. Since  the current treatment
 goes rather quickly  over some of the  steps,   the reader is referred
to the above citation for a more detailed discussion.
\par
The initiating point  of our formalism is 2+1-dimensional anti-de Sitter
gravity. Along with the classical action, we include a matter action
that  describes minimally coupled, massless, quantized  scalar fields. 
The complete action functional (up to surface terms)  can be written as
follows:
\be
I^{(3)}={1\over 16\pi G^{(3)}}\int d^{3}x\sqrt{-g^{(3)}} 
(R^{(3)}+{2\over l^2})  
 -{{\cal K} \over 16\pi G^{(3)}}\int d^{3}x\sqrt{-g^{(3)}}
\sum_{i=1}^{N}\left( \nabla^{(3)}f_{i}\right)^2,  
\label{1}
\ee
where $f_i$ denotes the matter fields, $N$ is a large  integer,
 $G^{(3)}$ is the 3-d Newton constant, $-2l^{-2}$ is the
negative cosmological constant and ${\cal K}$ is a  
 coupling parameter
that  vanishes in the classical limit (i.e., as $\hbar\rightarrow 0$). 
\par
Axial symmetry can be imposed on this action  by way of the following
metric ansatz \cite{21}:
\be
ds^{2(3)}=g_{\mu\nu}dx^{\mu}dx^{\nu}+\phi^2
(\alpha d\theta+ A_{\mu}dx^{\mu})^2,
\label{2}
\ee
where $\mu,\nu=\lbrace 0,1\rbrace$, $\alpha$ is an arbitrary constant of
dimension length, $A_{\mu}$ is a vector gauge field, $\phi$ is
 a scalar field (the ``dilaton'') and all fields are functions of only
$ \lbrace  x^0,x^1\rbrace  = \lbrace t,x \rbrace$. 
This reduction process results in the following 1+1-dilaton model:
\be
I= \int d^{2}x\sqrt{-g}\phi(R+2l^{-2}-{1\over 4}\phi^2
F^{\mu\nu}F_{\mu\nu})
-{\cal K} \int d^{2}x\sqrt{-g}\phi
\sum_{i=1}^{N}\left( \nabla f_{i}\right)^2,  
\label{3}
\ee
where we have set $\alpha=8G^{(3)}$   without loss of generality. The 
  ``field-strength'' tensor $ F_{\mu\nu}=\partial_{\mu}A_{\nu}-
\partial_{\nu}A_{\mu} $  is known to be
directly related to  the angular momentum of
   a  rotating  BTZ  black hole \cite{21}. 
Note that  the reduced action  describes  constant curvature gravity  with
coupling 
to both an Abelian gauge field and conformally invariant matter fields.
\par
Since  the action is ultimately significant  as the exponent in 
a path integral, it is possible
to ``integrate out'' the matter fields and then consider the vacuum 
limit (i.e., $f_i\rightarrow 0$). In this event, the 
resultant ``effective action''  can be
(at least) partially derived  in conformally invariant matter theories
because of its
exploitable  relation to the trace conformal anomaly \cite{27}.
For  the special case of conformally invariant matter in
 two spacetime dimensions,
the    effective action  can  be derived exactly up to terms that are
conformally invariant
\cite{28}.  For the dilaton model of interest, we find (assuming  
that $N>>1$ and the black hole is massive when compared  to the  Planck scale):
\be
I= I_{CL}
-{\cal K}\int d^2x\sqrt{-g}\left[R{1\over \Box}R
-{3\over\phi^2}(\nabla\phi)^2 \left({1\over \Box}R-\ln \mu^2\right)
-6\ln (\phi) R\right],
\label{4}\ee
where $I_{CL}$ is the left-most integral in Eq.(\ref{3}),  ${\cal K}$
has been appropriately rescaled (now,   ${\cal K}\approx N\hbar$)
and $\mu$ is an arbitrary parameter that arises out of renormalization
procedures \cite{27}.
 The precise forms of the  functional coefficients (in this case,  
$3/\phi^2$ and 
$6\ln \phi$) 
depend  upon the form of  dilaton-matter coupling that arises out of
the reduction process.
\par
It should be pointed out that the conformally invariant portion of 
the  effective action, which is described by the $\ln \mu^2$-term in 
Eq.(\ref{4}), is incomplete as shown. This portion cannot be found in
a closed form, but it can be approximated by an expansion (in powers
of curvature) of which we have only  included  the leading-order
 term \cite{NO2}.
Recently,   non-local terms of this expansion, which appear
to be
 relevant to the perturbative order of Eq.(\ref{4}), have been calculated 
\cite{ZEL}. Because of their non-local nature, the incorporation of  such
   terms into our formalism 
 is by no means a  straightforward process.  Consequently,
 for the sake of simplicity,  we have  
omitted  these terms  in the current analysis.  This issue is further
addressed  in the
final section.
\par
It is convenient to re-express the effective action  in an equivalent
 local form  as follows:
\bea
  I&=&I_{CL} -{\cal K}\int d^2x\sqrt{-g}
\left[(\psi + \chi) R +g^{\mu\nu}
\nabla_{\mu}\psi\nabla_{\nu}\chi 
\right.\nonumber\\
&&\quad\quad\quad\quad\quad\quad\quad\left.
-{3\over\phi^2}(\nabla\phi)^2\left(\psi-\ln \mu^{2}\right)
-6\ln (\phi) R \right],
\label{5}
\eea
where $\psi$ and $\chi$ are a pair of auxiliary scalar 
fields\footnote{Auxiliary fields of an analogous  form were first used 
in ref.\cite{BUR}
in the context  of  spherically symmetric gravity.} 
 that are
 constrained according to 
 the following  equations:
\be
\Box\psi=R,\label{6}\ee
\be
\Box\chi=R-{3\over\phi^2}(\nabla\phi)^{2}.\label{7}\ee
\par 
By varying
 the effective action (\ref{5}) 
with respect to the metric, dilaton and 
Abelian gauge field, we  obtain:
\be
-2\nabla_{\mu}\nabla_{\nu}\phi+2g_{\mu\nu}\Box
\phi-{2\over l^{2}}g_{\mu\nu}\phi
-{1\over 4}\left(3g_{\mu\nu}F^{\alpha\beta}F_{\alpha\beta}
-4g^{\alpha\beta}
F_{\mu\alpha}F_{\nu\beta}\right)\phi^3=T_{\mu\nu}
,\label{8}
\ee
\be
R+{2 \over l^{2}}-{3\over 4}\phi^2 F^{\mu\nu}F_{\mu\nu}
=D
,\label{9}
\ee
\be 
\nabla_{\mu}\left(F^{\mu\nu}\phi^3
\right)=0
\label{10}
\ee
(respectively), where:
\bea
T_{\mu\nu }&\equiv& -{\cal K} \biggl[2 \nabla_{\mu }\nabla_{\nu
}\left(\psi-6\ln (\phi)+\chi\right)  - \left(\nabla_{\mu}\chi\nabla_{\nu}\psi+
\nabla_{\mu }\psi\nabla_{\nu }\chi\right)\biggr.\nonumber\\
&&\quad\quad-g_{\mu\nu }\left(
2\Box\left(\psi-6 \ln (\phi)+\chi\right)
-g^{\alpha\beta}\nabla_{\alpha} \chi \nabla_{\beta}
\psi\right) 
\nonumber\\ &&\quad\quad\biggl.  
+{3\over\phi^2}\left(\psi-\ln \mu^{2}\right)\left(
2\nabla_{\mu}\phi\nabla_{\nu}
\phi- g_{\mu\nu}
(\nabla\phi)^2\right)\biggr]
,\label{11}
\eea
\be 
D\equiv 6{\cal K}\left[\phi^{-3}(\nabla\phi)^2(\psi-\ln \mu^2)+g^{\mu\nu}
\nabla_{\mu}\left(\phi^{-2}(\psi-\ln \mu^2)\nabla_{\nu}\phi\right)
-\phi^{-1}R\right].
\label{12}\ee
Note  that $T_{\mu\nu}$ can be identified with  the quantum stress tensor. 
\par
  The ``Maxwell'' field equation (\ref{10}) can be trivially integrated
to  yield: 
\be
F^{\mu\nu}={1\over l}{\epsilon^{\mu\nu}\over \sqrt{-g}}{J\over\phi^3},
\label{13}\ee
where $\epsilon^{\mu\nu}$ is the  Levi-Civita symbol
and $J$ is an integration constant that can be identified with the Abelian
charge observable (i.e., quantized angular momentum).  The above result
inspires the definition of an   ``effective  potential'' 
$V_J(\phi)\equiv l^{-2}(2\phi-\half J^2\phi^{-3})$,
which leads to the remaining field equations (\ref{8},\ref{9}) taking
on the following compact  forms:
\be
-2\nabla_{\mu}\nabla_{\nu}\phi+2g_{\mu\nu}\Box\phi-g_{\mu\nu}V_J(\phi)
=T_{\mu\nu},\label{14}\ee
\be
R+{\partial V_J\over\partial \phi}=D.
\label{15}\ee
\par
It is instructive to first consider the classical (${\cal K}=0$) solution. 
A prior work  has  demonstrated how to obtain the classical solution in
a static gauge  for a wide class
of    dilaton models \cite{30}. For  reduced BTZ gravity, this
solution can be expressed  as follows:
\be
\phi_{CL}={x\over l}
,\label{17}\ee
\be
(A_t)_{CL}=-{l^2 J\over 2 x^2}
,\label{AT}
\ee
\be
ds^2=-g_{CL}(x)dt^2+g^{-1}_{CL}(x)dx^2
,\label{18}\ee
\be
g_{CL}(x)={x^2\over  l^2}-lM + {J^2 l^2\over 4 x^2},
\label{19}\ee
where we have  assumed (without loss of generality) a timelike gauge vector
 and
 $M$ is a constant parameter that can be identified with
  the ADM  mass
observable. It is useful to note that
$R_{CL}=-g_{CL}^{\prime\prime}$ 
(where   primes indicate differentiation with respect
to $x$)  and  $g_{CL}(x)=-|k^{\mu}|^2$, where   
$k^{\mu}=l(\sqrt{-g})^{-1}\epsilon^{\mu\nu}\partial_{\nu}\phi$
 is a  Killing vector for the classical field equations.
\par
For subsequent calculations, it is convenient to re-express $g_{CL}(x)$ 
in the following form:
\be
g_{CL}(x)={1\over l^2 x^2}(x^2-x_{o}^2)(x^2-x_{i}^2),
\label{21}\ee
where:
\be
x_{o}^2={l^3\over 2}\left[M+\sqrt{M^2-J^2/l^2}\right],
\label{22}\ee
\be
x_{i}^2={l^3\over 2}\left[M-\sqrt{M^2-J^2/l^2}\right].
\label{23}\ee
The positive root  of $x^2_{o}$/$x^2_{i}$  locates  the classical outer/inner 
   event horizon. Since we have  restricted considerations to  black
hole solutions (and non-extremal ones  until Section 4), the phase space
of observables is restricted by $M>0$ and  $M^2>J^2/l^2$.
\par
For a higher-order analysis, it is necessary to introduce a
suitable ansatz for describing  the  back-reaction effects 
 on the 
classical geometry. Following a proposal by Frolov et al. 
\cite{32}, we now express the quantum-corrected solution
in the following manner:
\be
\phi=\phi_{CL}= x/l
,\label{24}\ee
\be
ds^2=-e^{2\omega(x)}g(x)dt^2+g^{-1}(x)dx^2
,\label{25}\ee
\be
g(x)=g_{CL}(x)-lm(x),
\label{26}\ee
where the fields $m(x)$ and $\omega(x)$ must vanish as ${\cal K}\rightarrow 0$.
Note that  $A_t=(A_t)_{CL}$ follows trivially, since we have assumed  
no coupling between the  matter and Abelian sectors.
\par
By substituting  the above ansatz into the  field  equations,
we find that Eq.(\ref{15}) and the off-diagonal component  of
Eq.(\ref{14}) are both  identically vanishing.    
After some simplification, the ``surviving'' field  equations  are found to be:
\be
-e^{2\omega}g m^{\prime}=T_{tt}
,\label{28}\ee
\be
-{m^{\prime}\over g}+{2\over l}\omega^{\prime}=T_{xx}.
\label{29}\ee
If  these expressions 
 are truncated  at the one-loop level (i.e., at first order in
${\cal K}$), then  we obtain the 
elegant
results:
\be
m^{\prime}=-T^{t}_{t}
,\label{30}\ee
\be
\omega^{\prime}={l\over 2 g_{CL}}(T^{x}_{x}-T^{t}_{t}).
\label{31}\ee
\par
Next in this  study, we  explicitly   formulate   the 
auxiliary fields $\psi(x)$ and $\chi(x)$. Since we are ultimately  deriving
one-loop expressions,  it is sufficient to express these fields in terms of
the classical  geometry. Furthermore,
the choice of boundary conditions should
reflect the Hartle-Hawking vacuum state \cite{24}.  Such conditions
restrict the analysis  to solutions that are   
 periodic in Euclidean (i.e., imaginary) time  when
on a spatial manifold extending  from the   outer horizon 
to a fixed outer boundary $L$
 \cite{6}.  
  \par
Let us first  consider solving  
  Eq.(\ref{6}) for $\psi$. The appropriate  solution  
 can be found
 by way of a special  map \cite{32}:  the classical Euclidean
geometry (in a static gauge) conformally mapped to the geometry 
 of a
   ``disc''. Significantly, the disc geometry  can be interpreted as  the
   Rindler coordinate description of 
  the
 Hartle-Hawking  state for a flat spacetime  \cite{4.50}. 
On the basis of
Eqs.(\ref{6},\ref{18}), 
such a   
map can be  suitably described by: 
\be
g_{CL}(x)(idt)^2+g_{CL}^{-1}(x)dx^2=e^{-\psi(z)}\left[z^2d\theta^2+dz^2\right],
\label{32}\ee
where the  disc coordinates are  confined  
to $0\leq \theta\leq 2\pi$ and $0\leq z\leq L_z$.
Solving for
  $\psi(z(x))$, we find:
\be
\psi (x) = -\ln g_{CL}(x) -{4\pi\over \beta_{CL} }\int^L_x{dx\over g_{CL}(x)}
-2\ln \left({\beta_{CL}\over 2\pi \zz}\right),
\label{33}
\ee
where  $\beta_{CL}$ denotes the Euclidean  time periodicity for
 the classical 
system
(i.e., 
$0\leq  it \leq\beta_{CL}$).
\par
We can
 determine  $\chi(x)$   by integrating  
Eq.(\ref{7}) and  then  imposing  the constraint that
 $\chi\rightarrow\psi$ in the limit of minimal dilaton-matter
coupling (for which the effective action assumes a ``Polyakov-like''
form \cite{33}). This procedure  leads to:\footnote{The
  $x=x_o$ integration limit (besides being an  intuitive choice) 
can be uniquely  fixed  by constraining 
 the curvature $R$  to be  regular on the horizon.}   
\be
\chi(x)=\psi(x)+3\int^L_x {dx\over
g_{CL}(x)}\int^x_{x_{o}}{d{\tilde {x}}
g_{CL}({\tilde{x}})\over {\tilde{x}}^2}.
\label{34}\ee
\par
By substituting the classical solution 
(\ref{17},\ref{18},\ref{33},\ref{34}) into  the stress tensor  (\ref{11}),
we obtain  the following one-loop expressions:

\bea
T^t_t  &=&{{\cal K}\over  g_{CL}}\left[
(g_{CL}^{\prime})^2-
 4g_{CL} g_{CL}^
{\prime\prime}-{16\pi^{2} \over \beta_{CL}^2}
+6 { g_{CL}\over x^2}\left(2g_{CL}-xg_{CL}^{\prime}\right)
\right. \nonumber \\ && \left. - 3
{g_{CL}^2\over x^2}\left( 2+ \ln  g_{CL} +{4\pi \over\beta_{CL}}\int^L_x 
{dx\over g_{CL}}+\ln \Upsilon^2\right) 
 +{12\pi\over\beta_{CL}}
\int^x_{x_{0}}  {dx g_{CL}\over x^2}\right]
,\label{35}
\eea
\bea
T^x_x &=&{{\cal K} \over  g_{CL}}\left[{16\pi^{2}
\over\beta_{CL}^{2}}-(g_{CL}^{\prime})^2
-{6\over x}g_{CL}g_{CL}^{\prime}\right.\nonumber\\
&&\left.
+ 3{g_{CL}^2\over x^2}\left(  \ln  g_{CL}+{4\pi\over\beta_{CL}}
\int^L_x {dx\over g_{CL}}+\ln \Upsilon^2\right)
-{12\pi\over\beta_{CL}}
 \int^x_{x_{0}}{ dx
g_{CL}\over x^2}\right],  
\label{36}
\eea
where $\Upsilon\equiv\mu\beta_{CL}/2\pi L_{z}$  can be regarded as
 an arbitrary parameter.
\par
When on the  constraint surface,
 the Euclidean time periodicity   must be suitably  fixed  
 to 
  ensure the horizon  regularity of  the Euclidean  geometry 
(i.e.,   to  eliminate  any  conical singularity or   deficit angle) 
 \cite{6}.  
On this basis, we can explicitly evaluate the on-shell value of
$\beta_{CL}$ by  matching the classical solution with a conical geometry:
\be
e^{2\omega(x)}g(x)(idt)^2+g^{-1}(x)dx^2=z^2d\theta^2+H(z)dz^2 
\label{49A} \ee
(where $0\leq\theta\leq2\pi$ and $z=0$ at $x=x_o$) and then
enforcing   $H(0)=1$. This process yields:
\be
\beta_{CL}=\left.{4\pi\over g_{CL}^{\prime}}\right|_{x=x_o}= {2\pi l^2\oox\over
x_o^2-x_i^2}.  
\label{50A}\ee
\par
 By substituting Eq.(\ref{35}) into Eq.(\ref{30}), integrating
 and also incorporating   Eqs.(\ref{21},\ref{50A}), we find:
\bea
m(x)&=& 2{{\cal K}\over l^2}\biggl[2x-3
{\oox^2+\iix^2\over  x}-8\iix
\ln \left({x-\iix\over x+\iix}\right)\biggr.
\nonumber\\&&\quad\quad+\half\left(3x+3{\oox^2+\iix^2\over x}-{\oox^2\iix^2
\over x^3}\right)
\biggl(
{\iix\over\oox}\ln \left({x-\iix\over x+\iix}\right)\biggr.\nonumber\\ &&
\quad\quad\quad\quad\quad\quad\biggl.\biggl.
+\ln \left({(x+\oox)^2(x^2-\iix^2)\over l^2 x^2}\right)+\Theta
\biggr)\biggr]+m_0,\label{37}
\eea
where:
\be
\Theta\equiv -{\iix\over\oox}\ln \left({\LLx-\iix\over L+\iix}
\right)+\ln \left({\LLx-\oox\over L+\oox}
\right)+\ln \Upsilon^2\label{38}\ee
and $m_0$ is an integration constant that can be absorbed (without loss
of generality)   into the classical
mass $M$.  Next, let us invoke the convention   $\omega(L)=0$ and define
a function  $\varpi(x)$  in accordance with 
$\omega(x)={\cal K} l \left(\varpi(L)-\varpi(x)\right)$.
 Then the substitution of
Eqs.(\ref{35},\ref{36}) into  Eq.(\ref{31}) ultimately yields:
\bea
\varpi(x)&=&- {1\over x}+2{3\oox^2+\iix^2\over (\oox^2-\iix^2)(x+\oox)}
-2{\iix^2(\oox^2+3\iix^2)+\oox(\oox^2-5\iix^2)x\over \oox(\oox^2-\iix^2)
(x^2-\iix^2)}
\nonumber\\&&\quad
-8{\iix\over(\oox+\iix)^2}\ln \left({x-\iix\over x+\oox}\right)
+8{\iix\over(\oox-\iix)^2}\ln \left({x+\iix\over x+\oox}\right)
\nonumber\\&&\quad +{3\over x}\biggl[
{\iix\over\oox}\ln \left({x-\iix\over x+\iix}\right)
+\ln \left({(x+\oox)^2(x^2-\iix^2)\over l^2 x^2}\right)+\Theta
\biggr].
\label{552}\eea
Note  that
 $m(x)$ and $\omega(x)$ 
 are both well-defined functions  for $x_o\leq x \leq L$,
thereby substantiating our choice of ansatz.
\par
We next   consider the  quantum-corrected curvature. This can be written as 
$R=-e^{-\omega}(e^{-\omega}(e^{2\omega}g)^{\x})^{\x}$ or for a one-loop
truncation:
\be
R=-g_{CL}^{\prime\prime}+lm^{\prime\prime}-2\omega^{\prime\prime}g_{CL}
-3\omega^{\prime}g_{CL}^{\prime}.\ee
 Substituting  the prior results for $m(x)$ and $\omega(x)$,  evaluating
the derivatives
 and then   simplifying, we obtain:
\bea
R&=& -{2\over x^4 l^2}(x^4+3\oox^2\iix^2)+{2{\cal K}\over l x}
\biggl[24 + {6\over \oox x}(\oox^2-\iix^2)\biggr.\nonumber\\
 &&+{4\over \oox x^4 (x+\oox)^2 (x^2-\iix^2)^2}\biggl(
\iix^6\oox^2(3x^3+6\oox x^2 +8\oox^2 x+ 4\oox^3)
\biggr.\nonumber\\&&
-\iix^4 x^2( 3x^5+6\oox x^4 +5 \oox^2 x^3+8\oox^3 x^2+ 11\oox^4 x+6\oox^5)
\nonumber\\&&\biggl.
-\iix^2 x^5(3x^4-4\oox x^3-11\oox^2 x^2 -6\oox^3 x+3\oox^4)
 +3 x_o^2 x^9\biggr)\nonumber\\ &&
+{3\over x^4}\left(3x^4-(\oox^2+\iix^2)x^2 -\oox^2\iix^2\right)\times
\nonumber\\
&&\biggl.\quad\quad\quad \biggl(
 {\iix\over\oox}\ln \left({x-\iix\over x+\iix}\right)
+\ln \left({(x+\oox)^2(x^2-\iix^2)\over   l^2 x^2}
\right)+\Theta\biggr)
\biggr],\eea
which is also a  well-defined quantity  throughout the relevant manifold.
\par
Let us next consider 
the one-loop shift in the outer horizon $\Delta x_o$.  
 To determine this  shift, we  begin  with a first-order   Taylor expansion of 
 the function $g(x_o+\Delta x_o)$; cf. Eq.(\ref{26}).
 After expanding and  applying  the horizon conditions
$g_{CL}(x_o)=g(x_o+\Delta x_o)=0$ (with the latter valid being
valid to first order), we find:
\be
\Delta x_{o}={l^3 x_o\over 2 (x_o^2-x_i^2)}m(\oox)={l \beta_{CL} \over
4\pi}m(\oox).
\label{43}\ee
Note that
 a similar  calculation is  not viable at  the inner horizon, 
since the  back-reaction ansatz
 has  not been  strictly  defined for $x<x_o$. Furthermore,  
the  shift in $x_i$  is
expected to be  non-analytic in  ${\cal K}$ \cite{27}.

\section{Non-Extremal Thermodynamics}

 Our method of thermodynamic analysis is based on the well-known techniques
of  Gibbons,  Hawking \cite {6} and others \cite{25,32}.
 This procedure can be summarized as follows. After  analytically 
continuing  to 
Euclidean spacetime and    closing off the imaginary   time direction,
one finds that
 the path integral can  be  
interpreted as a thermodynamic partition function $\cal{Z}$.
This partition function describes
an ensemble of black holes that are radiating at a temperature 
 $\beta^{-1}$, where  $\beta$  corresponds to   the periodicity of
 Euclidean time.  
Furthermore, a semi-classical approximation has been shown to
yield  the relation \cite{6}:
\be 
\ln ({\cal Z})=-I_{OS}, \label{PART}\ee
where $I_{OS}$ denotes the on-shell Euclidean action.
Note for an on-shell system that $\beta^{-1}$ corresponds to the
so-called ``Hawking temperature'' of black hole radiation.\footnote{
Keep in mind that an observer at $x$ locally  measures an inverse 
temperature of $\sqrt{-g_{tt}[x]}\beta$; that is,  a ``red-shifted''
 value of inverse temperature
 \cite{25}.  For anti-de Sitter spacetimes (unlike for
 asymptotically flat ones),
this red-shift factor diverges as $x\rightarrow\infty$.}  
\par
Let us reconsider the effective action of  Eq.(\ref{5}).
By transforming to  Euclidean spacetime   
(i.e., rotating  $t\rightarrow it$ and 
 re-expressing all  geometrical objects   in terms of a  positive-definite
metric\footnote{Technically, the  Abelian charge should also be complexified
so that $A_{t}dt$ remains invariant \cite{25}.  It is implied, however, that
 we have already  continued back to real charge before presenting  any
result in this paper.})
 and  also  applying the  static solution of  Eqs.(\ref{24},\ref{25}),
we obtain the following Euclidean form of the action:
\bea
I&=&-\beta\int^L_{x_q} dx e^{\omega}\biggl[ {x\over l}R +V_J({x\over l})
-{\cal K} \biggl((\psi-6\ln ({x\over l})+\chi)R
\biggr.\biggr.\nonumber\\ && 
\quad\quad\quad\quad\quad\quad\quad\quad\quad\quad\quad\quad\quad\quad
\biggl.\biggl.
+g\chi^{\prime}\psi^{\prime}
-3{g\over x^2}(\psi-\ln \mu^2)\biggr)\biggr] 
\nonumber\\&&
+\beta\left.\left(\int^L dx e^{\omega}R\right)
\left[{x\over l}-{\cal K}\left(\psi-6\ln ({x\over l})+\chi\right)
\right]\right|_{x=L}-{\beta J\over l}\Delta A_t, 
\label{45}\eea
where $x_{q}$  represents  the quantum-corrected outer horizon, 
 $\Delta A_t\equiv[A_t(L)-A_t(x_q)]$ and note that $e^{\omega}R$ is a total 
derivative. So as 
to ensure a well-defined variational principle at the boundaries of the system,
we have included the appropriate surface terms  in the third line
of this  expression.      
Except for the  right-most (charge sector)   term,
this surface contribution  is directly analogous to
Gibbons and Hawking's   ``extrinsic 
curvature term'' \cite{6}.\footnote{
 Technically, we should also  include 
  an analogous horizon term, as well as
a delta-function contribution  from the curvature  \cite{35}. 
However,  these horizon  contributions are known to  cancel off,  
provided a regularized conical singularity \cite{32}. 
Hence, these terms   are not pertinent to  
on-shell thermodynamics.}
\par
The Euclidean  action can be written  in a more convenient form  by
way of  Eq.(\ref{14}).  Let us first define $G_{\mu\nu}$ as the left-hand side
of this field equation and then   express both  $G_{tt}$ and $T_{tt}$ 
(which can be obtained  from Eq.(\ref{11})\footnote{It is helpful  
 to  first  make the substitution 
 $\Box(\psi+\chi)=\Box(2\psi-3\phi^{-2}
(\nabla\phi)^2)$; cf. Eqs.(\ref{6},\ref{7}).}) in terms of the static
solution. After integrating the curvature terms in the 
    Euclidean action (\ref{45}) by parts, we can incorporate the static
forms of   $G_{tt}$ and $T_{tt}$ 
 to obtain: 
\bea
I&=&
 \beta \int^L_{x_q} dx \biggl[  
e^{\omega}(G^t_t-T^t_t) \biggr.
\nonumber\\ 
&&\quad\quad\quad\quad \biggl. -\biggl(
{2\over l}e^{\omega}g
+4{\cal K} e^{-\omega}(e^{2\omega}g)^{\x} +2{\cal K}
e^{\omega}g\left(
\psi+6\ln ({x\over l})-\chi\right)^{\x}\biggr)^{\x}\biggr]
\nonumber\\&&
+\beta\left.\left(\int^{x_q} dx e^{\omega}R\right)
\left[{x\over l}-{\cal K}\left(\psi-6\ln ({x\over l})+\chi\right)
\right]\right|_{x=x_q}-{\beta J\over l}\Delta A_t. 
\label{48}\eea
Evidently,   the 
above  integrand  vanishes  on the constraint  surface up to
a total divergence.  It follows that
  the on-shell  Euclidean action  reduces to just a 
surface expression, and this is found to be:
\bea
I_{OS}&=&-\beta\left.\left[{2\over l}g +4{\cal K}(e^{2\omega}g)^{\x}
+2{\cal K} g\left(\psi+6\ln ({x\over l})-\chi\right)^{\x}\right]\right|_{x=L}
\nonumber\\ && 
-4\pi\left.\left[{x\over l}-{\cal K}\left(\psi-6\ln ({x\over l})+\chi\right)
\right]\right|_{x=x_q}-{\beta J \over l} \Delta A_{t}.
\label{51}\eea
Here, we have  used $\omega(L)=g(x_q)=0$ and the perturbative analogue of  
Eq.(\ref{50A}):
\be
\beta=4\pi \left.{e^{-\omega}\over g^{\prime}}\right|_{x=x_{q}}.
\label{50}\ee
\par
We now recall the relation $\ln ({\cal Z})=-I_{OS}$  and  point out  that
(on the basis of thermodynamic arguments) 
the logarithm of the partition function  should ultimately take on 
the following  free energy form: 
\be
\ln ({\cal Z})=-\beta_L\left[E-\sum_{\eta}\eta\gamma_{\eta}\right]+S,
\label{52}\ee
where $\beta_L$ is the  fixed value of 
 inverse temperature at the outer boundary of the system,
 $E$ is the thermal   energy, $\eta$ is an intrinsically  conserved quantity,
$\gamma_{\eta}$ is the related chemical potential and
$S$ is the entropy of the system. 
By comparing the two expressions for $\ln {\cal Z}$, we are able to
make the following identifications:
\be
E=-2{\beta\over\beta_{L}}\left.\left[ {1\over l}g_{CL}-m
+2{\cal K} g_{CL}^{\x}
+{\cal K} g_{CL}\left(\psi+6\ln ({x\over l})-\chi\right)^{\x}\right]
\right|_{x=L}
,\label{53}\ee
\be
S={4\pi\over l}(x_o+\Delta x_o)-4\pi{\cal K}\left.
\left[\psi-6\ln ({x\over l})+\chi\right]\right|_{x=x_o}
,\label{54}\ee
\be
\gamma_J={1\over l}{\beta\over\beta_{L}}\left[A_t(L)-A_t(x_o+\Delta x_o)
\right].
\label{55}\ee
\par
Before any  further evaluation, two  points should  be clarified:
\\ (i) The inverse boundary   temperature $\beta_L$ is ``red shifted''
from  the inverse Hawking  temperature  $\beta$
according to \cite{25}
$\beta_L= \sqrt{g_{tt}[x=L]}\beta=\sqrt{g(L)}\beta$. \\ 
 (ii)  The Euclidean  action  is known to    diverge
as the outer boundary tends to infinity \cite{6}, which  implies that the
calculated  energy will also  diverge   unless a suitable subtraction
 procedure is
 invoked.
The usual convention is to
subtract off the energy contribution from   the asymptotic geometry
\cite{25}, and so we define a subtracted energy  according to 
 $E_{sub}\equiv E\left[g(L)\right]-E\left[g_{\infty}\right]$,
where $g_{\infty}\equiv L^2/l^2$. 
\par
By
substituting the prior geometrical  formalism into 
Eqs.(\ref{50},\ref{53}-\ref{55}) and also
using binomial expansions where applicable, we obtain the following one-loop
expressions:
\bea
T\equiv\beta^{-1}=\beta_{CL}^{-1}
&+&{\cal K} l\beta_{CL}^{-1}
\biggl[-{9\oox^4+6\oox^2\iix^2
+\iix^4\over(\oox^2-\iix^2)^2\oox}\biggr.\nonumber\\&&\biggl.
+16{\oox\iix^2\over(\oox^2-\iix^2)^2}\biggl(2 \ln \left({l \over \oox}
\right)+\Theta\biggr)+\varpi(L)\biggr]
,\label{59}\eea
\bea 
E_{sub}=
2{L\over l^2 }
\left(1-{1\over L^2}\sqrt{(\LLx^2-\oox^2)(\LLx^2-\iix^2
)}\right)
+{l L m(\LLx)\over
\sqrt{(\LLx^2-\oox^2)(\LLx^2-\iix^2)}}\nonumber\\
+2{{\cal K} \over l }\biggl[13 -{3\oox\over L}
-{13\oox L^4-2(3\oox^2+\iix^2)L^3
-3\oox(\oox^2+\iix^2)L^2+\oox^3\iix^2\over
\oox\LLx^2\sqrt{(\LLx^2-\oox^2)
(\LLx^2-\iix^2)}}\biggr]
,\label{60}\eea
\bea
S &=&    {4\pi\oox\over l}
 - {4\pi{\cal K}\over\oox^2-\iix^2}
\biggl[\oox^2+3\iix^2
+{\iix\over\oox}(3\oox^2+\iix^2) \ln \left({L-\iix\over L+\iix}\right)
\biggr.\nonumber\\
&&+2(3\oox^2+\iix^2)\ln \left({L+\oox\over \oox}\right)
-(\oox^2+3\iix^2)\ln \left({L^2-\iix^2\over\oox^2-\iix^2}\right)
\nonumber\\&&\biggl.-(\oox^2-\iix^2)\ln \left({\oox^5 L\over l^6}\right)
-(5\oox^2-\iix^2)\biggl(\ln \left({\oox^2-\iix^2\over l^2}\right)+
\Theta\biggr)
\biggr],\label{61}\eea
\be
\gamma_J={l^2J(\LLx^2-\oox^2)\over 2\oox^2 L\sqrt{(\LLx^2-\oox^2)
(\LLx^2-\iix^2)}}
\left[1+\half{ l^3 L^2\over L^2-\oox^2}
\left({ m(\LLx)\over\LLx^2-\iix^2}-
{2 m(\oox)\over \oox^2-\iix^2}\right)\right].
\label{62}\ee
\par
A brief comment regarding the one-loop
 entropy is in order. Although the  black hole entropy is
 normally a property of
the horizon, the above expression (\ref{61}) contains terms that depend
on the ``box size'' $L$.
 This paradoxical behavior  can be attributed
to the non-local nature of the auxiliary fields $\psi$ and $\chi$; 
cf. Eqs.(\ref{6},\ref{7}).
 Even 
 at the horizon, these fields  contain information with regard to the 
entire manifold.
Physically, the  $L$-dependent terms  can be   attributed to 
 a ``hot  thermal gas'' that  fills up  the box.
\par
For a check on validity, it is helpful to consider the classical  limit.
Firstly, we can re-express the classical entropy  
 in  the usual    ``Bekenstein-Hawking'' form (i.e., $S=A/4G^{(3)}$)  by
making the following identification   (cf. Eq.(\ref{2})):
 $A=2\pi\cdot 8G^{(3)}\phi(x_o)$  is 
 the circumference of the outer horizon. 
Let us  next  consider the   behavior of the classical energy in the
$L\rightarrow\infty$ limit. Under these conditions, it can be shown that
  $\sqrt{g(L)}E_{sub}\rightarrow M$, 
which   is    
 the expected  asymptotic behavior of a quasi-localized energy in
  an anti-de Sitter spacetime \cite{50}.
 A similar analysis  for the  
 chemical potential yields the limit
  $\sqrt{g(L)}\gamma_{J}\rightarrow lJ/2 x_o^2$, 
 which is the form of rotational potential that might  be
anticipated 
for an axially symmetric system of radius $x_o$ and angular momentum $J$.
 Finally, it  can be   readily
verified \cite{4.50}
 that the  classical limit of $T$
 satisfies the expected  relation between the Hawking temperature
and  the surface gravity; i.e.,
$T=\kappa/2\pi$.
\par
A final thermodynamic consideration is the flux of thermal radiation.
This flux has both an emission and absorption  component  that are equal
in magnitude (assuming  the Hartle-Hawking state).  Furthermore,
it has been shown \cite{39} that the flux components  are equivalent
to the diagonal components
of the stress tensor  if  these tensor components are
 expressed in terms of   
 suitably defined null  coordinates.  
In regard to the
classical BTZ  geometry, the appropriate  coordinates can be
 defined as follows:
\be
u=t-\int{dx\over g_{CL}},\quad v=t+\int{dx\over g_{CL}}.
\label{66}\ee
  It  can be  readily shown that:
\be
T_{uu}=T_{vv}=-{g_{CL}\over 4}(T^t_t-T^x_x).\label{68}\ee
Note that $T_{uu}$/$T_{vv}$ represents  the outgoing/incoming flux
and   $T_{uv}$ can be   obtained by ``flipping'' the sign in front of $T^x_x$.
\par
By incorporating    Eqs.(\ref{35},\ref{36}) into the above relation,
 we  find   the following results:
\bea
T_{uu}&=&- {{\cal K}\over 2}{ (x-\oox)^2\over   l^4\oox^2 x^6}\biggl[
3\oox^2 x^6 +6\oox(3\oox^2-2\iix^2)x^5
+(3\oox^4-2\oox^2\iix^2+4\iix^4)x^4\biggr.\nonumber\\&&\quad
+4\oox\iix^2(2\oox^2-\iix^2)x^3
+3\oox^2\iix^2(2\oox^2-3\iix^2)x^2
-10\oox^3\iix^4 x-5\oox^4\iix^4
\nonumber\\&&\quad
-3\oox^2(x+\oox)^2(x^2-\iix^2)^2
\biggl(
{\iix\over\oox}\ln \left({x-\iix\over x+\iix}\right)\nonumber\biggr.\\
&&\quad\quad\quad\quad\quad\quad\quad\quad\quad\quad\quad\quad
\biggl.\biggl.
+\ln \left({(x+\oox)^2(x^2-\iix^2)\over l^2x^2}\right)+\Theta\biggr)\biggr]
,\label{70}\eea
\be
T_{uv}= {{\cal K}\over 2}{ (x^2-\oox^2)(x^2-\iix^2)\over  l^4  x^6}
\biggl[13x^4+3(\oox^2+\iix^2)x^2-3\oox^2\iix^2
\biggr].\label{71}\ee
Note   the
 divergence  of these components   as $x\rightarrow\infty$.
An  asymptotically  divergent flux  is an  expected  outcome  for  
 an anti-de Sitter theory.  Since an asymptotic observer
locally measures a vanishing temperature (see Footnote $\#$2), 
it follows that she would  detect an infinite  flux of particles.
\par
The above calculations provide a further check on our formalism.
Christensen and  Fulling \cite{34} have shown  that enforcing 
stress tensor regularity at the outer  horizon 
in the free-falling frame  (which is a necessary condition for describing  the
Hartle-Hawking state)
leads to a certain class of   constraints. 
 These  translate to  the
 (outer) horizon regularity  of the following three quantities:
\be
(i) T_{vv}, \quad  (ii) T_{uv}/g_{CL}, \quad  (iii) T_{uu}/g_{CL}^2.
\label{CF}\ee 
The above expressions   satisfy all three of these constraints
by virtue of  the  $x-x_o$ factor(s) in front.

\section{Extremal Limit}

It is straightforward to consider the extremal limit  of the prior
 calculations. By definition, the extremal  limit corresponds to a 
coincidence in the  classical horizons:\footnote{By 
invoking a limiting procedure,
it is implied that the extremal condition may  be
violated by radiation effects. That is, the one-loop corrected horizons
may or may not coincide.}    $x_{i}\rightarrow x_{o}$
  or  $J^2\rightarrow l^2M^2$. This limiting  procedure leads to the 
following  results: 
\bea
m(x)&=&2{{\cal K}\over l^2}
\biggl[2x-6{\oox^2\over x}-8\oox
\ln \left({x-\oox\over x+\oox}\right)
\biggr.\nonumber\\&&\quad\quad\biggl.
+\half
(3x+{6\oox^2\over x}-{\oox^4\over x^3})\biggl( 2 
\ln \left({x^2-\oox^2\over 
l x}\right)+ \ln \Upsilon^2\biggr)
\biggr],\label{75}\eea 
\be
\varpi(x)\rightarrow \quad quadratically \quad divergent \quad throughout \quad
 the \quad manifold
,\label{76}\ee
\bea
R&=&-2{x^4+3\oox^4\over l^2 x^4}+ {2{\cal K}\over lx}\biggl[24+4{\oox^2
\over x^4}\left(x^2+\oox^2\right)\biggr.\nonumber\\
&&\quad\quad\quad\quad\biggl.+{3\over x^4}(3x^2+\oox^2)(x^2-\oox^2)
\biggl(2\ln \left({x^2-\oox^2\over l x}\right)+\ln \Upsilon^2\biggr)\biggr]
,\label{77}\eea
\be
\Delta x_o \rightarrow \quad linearly \quad divergent
,\label{78}\ee
\be
T= 0\quad + \quad linearly \quad divergent \quad corrections
,\label{79}\ee
\be
 E_{sub}= 2{x_o^2\over l^2 L}+ {l L  m(\LLx)\over\LLx^2-\oox^2
}
+2{\cal K} {x_{o}\over l  L^2}\left({5L^2-2x_o L+x_o^2\over L+x_o}\right)
,\label{80}\ee
\be
S={4\pi x_o\over l}\quad+ \quad linearly \quad divergent \quad corrections
,\label{81}\ee
\be
\gamma_{J}={l^2 J\over 2 L x_o^2}\quad
+\quad linearly \quad divergent \quad corrections
,\label{82}\ee
\bea
T_{uu}&=&- {{\cal K}\over 2}{ (x-\oox)^2\over   l^4  x^6}
\biggl[3x^6+6\oox x^5+5\oox^2 x^4+4\oox^3 x^3-3\oox^4 x^2-10\oox^5 x
\biggr.\nonumber\\&&\quad\biggl. -5\oox^6
-3(x-\oox)^2(x+\oox)^4\biggl(
2\ln \left({x^2-\oox^2\over l x}\right)
+\ln \Upsilon^2
\biggr)\biggr],\label{83}\eea
\be
T_{uv}= {{\cal K}\over 2}{ (x-\oox)^2(x+\oox)^2\over  l^4  x^6}
\biggl[13x^4+6\oox^2 x^2-3\oox^4\biggr].
\label{84}\ee
\par
With only a few exceptions (energy, curvature and flux),
we find the one-loop results to be poorly defined in the extremal limit.
 (Note that $m(x)$
has a logarithmic divergence at the outer  horizon.) 
Furthermore,   the stress tensor component $T_{uu}$    fails  the
previously discussed 
 regularity condition (\ref{CF}),\footnote
{It has been argued \cite{14}  that the same conditions  apply to the 
extremal case,  in spite of the difficulties in    formalizing
an extremal analogue to 
 ``Kruskal-like'' (i.e., free-falling)
coordinates.}
since  we now have 
$g_{CL}\propto (x-\oox)^2$.  One must conclude  that the one-loop  ansatz 
breaks down
in this  extremal limiting case.

\section{Alternative Approach to Extremal Case}

In this section, we reconsider the extremal case by invoking an ansatz
(for  quantum corrections)
that presumes an extremal solution from the beginning. There is
ample justification for such a procedure  because of  topological
differences in the extremal and non-extremal solutions \cite{7}.
\par
The methodology of this section    is to repeat the prior calculations
with   three fundamental  differences: \\
(i) In place of the  classical metric function of  Eq.(\ref{21}),
we now use:
\be
g_{CL}(x)={1\over l^2x^2}(x^2-\oox^2)^2,
\label{85}\ee
where $x_o=\sqrt{l^3 M/2}$ and it is useful to remember $g_{CL}^{\x}(x_o)=0$. 
Note that  the perturbative  ansatz of
Eqs.(\ref{24}-\ref{26}) is   otherwise  unaltered.\\
(ii) We now regard the  Euclidean time periodicity $\beta$  as  an  
arbitrary quantity.   This proposal  is based on the
 following  observation: 
  the extremal (Euclidean) geometry has   no conical singularity to
be regulated \cite{5}. \\
(iii)
In solving for 
 the auxiliary fields $\psi$ and $\chi$, we employ a
 different  method of imposing Hartle-Hawking boundary conditions.    First, 
 the associated  field  equations (\ref{6},\ref{7}) can be
directly integrated  to yield:
\be
\psi(x)=-\ln g_{CL} +{C_{\psi}\over l^2}\int^x {dx\over g_{CL}} 
,\label{87}\ee
\be
\chi^{\x}=-{1\over g_{CL}}\left[g_{CL}^{\x}-{C_{\chi}\over l^2}
+3\int^x dx {g_{CL}\over x^2}\right],
\label{88}\ee
where $C_{\psi}$ and $C_{\chi}$ are integration constants of dimension length.
(Note  that  the second integration  constant in $\psi$ 
 can be  absorbed into
$\ln \mu^2$ without loss of generality, whereas $\chi$ only  appears in the  
formalism  as a derivative.) The next step in this method 
 is to constrain  the pair of
  integration  constants. For this purpose, we impose  (outer)   horizon
regularity on  three geometrical functions: $m(x)$, $\varpi(x)$ and $R(x)$.
By evaluating each of  these quantities (for arbitrary $C_{\psi,\chi}$)
and  locating  the horizon singularities in the resultant
 expressions, we are 
able to identify 
the  following set of  constraints:\\ 
(a) $m\rightarrow\quad C_{\psi}+C_{\chi}=8x_o\quad or\quad C_{\psi}=0$,\\
(b) $\varpi\rightarrow\quad C_{\psi}^2
-{3\over 2}C_{\psi}C_{\chi}-24C_{\psi}x_o
=-128x_o^2\quad and \quad (a)$, \\
(c) $R\rightarrow\quad 2C_{\psi}-3C_{\chi}=16x_o$; \\
which has  a  unique solution of:
\be
C_{\psi}=8x_o  \quad and \quad   C_{\chi}=0.
\label{92}\ee
\par
With regard  to  (iii),
it is worth noting that the same method can  be  applied  to the 
non-extremal case.  By  imposing  horizon regularity on the non-extremal 
geometry,  we find that
$C_{\psi}=2(x_o^2-x_i^2)/x_o$ and $C_{\chi}=2(3x_o^2+x_i^2)/x_o$; which
is  consistent  with the prior results  for $\psi$ and $\chi$
(\ref{33},\ref{34}). 
This is not surprising,
since the  specification of a quantum state (such as the 
Hartle-Hawking state)
should  uniquely determine these Green's  functions \cite{34}.
\par
With the   new   ansatz being  rigorously stipulated, we are now in a position
to re-evaluate the  extremal  black hole properties.  These results are
reported 
 below  with commentary wherever  clarity is required:
\bea
m(x)&=& m_0+2{{\cal K}\over l^2}\biggl[2x-6{x_o^2\over x}+2{x_o^3\over x^2}
-16{x_o^2\over x+x_o}\biggr.\nonumber\\ 
&& \biggl. \quad\quad\quad\quad
+(3x+6{x_o^2\over x}-{x_o^4\over x^3})\biggl(
\ln \left({(x+x_o)^2\over l x}\right)+\Theta\biggr)\biggr],
\label{94}\eea
where we have  redefined:
\be
\Theta\equiv \ln \left({L-x_o\over L+x_o}\right)
-2{x_o L\over L^2-x_o^2}+{\half}
\ln \mu^2
\label{95}\ee
and    $m_0$ is a constant that  must be constrained to satisfy  $m(x_o)=0$.
This constraint becomes  evident when we consider
a first-order Taylor expansion of  $g(x_o+\Delta x_o)$.
Note that no such method of fixing $m_0$  is apparent  in the 
 non-extremal analysis.
\par
\be
\varpi(x)=-{1\over x}+ { 4\over x+x_o} - 8{x_o\over (x+x_o)^2}
+{8\over 3} {x_o^2\over (x+x_o)^3} + {6\over x}\biggl(
\ln \left({(x+x_o)^2\over l x }\right)+\Theta\biggr)
,\label{96}\ee
\bea
R(x)&=&-2{x^4+3x_o^4\over l^2 x^4}+ 4{{\cal K}\over l x}
\biggl[12{x+x_o\over x}+6{x_o\over x^3}(3x^2+x_o^2)
\biggr.\nonumber\\
&&\quad\quad\quad\quad
+8{x_o^2\over x^4}
(x^2+x_o^2)
+ 8{x_o\over x}{(3x+x_o)(x^2+x_o^2)\over (x+x_o)^3}
\nonumber\\ &&\biggl.\quad\quad\quad\quad
+{3\over x^4}(3x^2+x_o^2)(x^2-x_o^2)
\biggl(\ln \left({(x+x_o)^2\over l x }\right)
+\Theta\biggr)\biggr]
,\label{97}\eea
\be
\Delta x_o=l\left.{m^{\x}\over g_{CL}^{\x\x}}\right|_{x=x_o}=2{\cal K} l.
\label{98}\ee
For this calculation, we have  considered a first-order Taylor expansion of
 $g^{\x}(x_o+\Delta x_o)$, since such an  expansion for  $g$ leaves 
$\Delta x_o$ as an indeterminate quantity. If we impose the one-loop constraint
  $g^{\x}(x_o+\Delta x_o)=0$
 (which is justified by
Hawking's conjecture \cite{5}:  an extremal black hole should
 retain
its   nature, regardless of  radiation effects) on the expansion in question, 
then  Eq.(\ref{98}) follows.
\par
\be
T=\beta^{-1}\rightarrow \quad indeterminate\quad (arbitrary\quad 
by \quad hypothesis)
,\label{99}\ee
\be
E_{sub}=2{x_o^2\over  l^2 L}+{lLm(L)\over L^2-x_o^2}
+2{{\cal K} x_o\over l L^2 (L^2-x_o^2)}\left[5L^3-7x_oL^2-5x_o^2L-x_o^3\right]
,\label{100}\ee
\be
S=0.
\label{101}\ee
This result of vanishing entropy occurs trivially, as    
the on-shell Euclidean action is now
linearly proportional to $\beta$. Furthermore, the horizon surface  term  
(which  normally
accounts 
for the entropy) must  vanish  according 
to $g(x_q)=g^{\x}(x_q)
=0$ (cf. Eq.(\ref{48})).
\par
\be
\gamma_J={l^2 J\over 2 L x_o^2}\biggl[1+\half{l^3L^2\over L^2-x_o^2}
\biggl({m(L)\over L^2-x_o^2}-{8{\cal K}\over l^2x_o}\biggr)\biggr]
,\label{102}\ee
\bea
T_{uu}&=&-{{\cal K}\over l^4}{(x^2-x_o^2)^4\over x^6}\biggl[
2{x_o\over (x+x_o)^4}\left(3x^3-2x_ox^2-5x_o^2x-2x_o^3\right)
\biggr.\nonumber\\
&&\biggl. \quad\quad\quad\quad\quad\quad\quad\quad
-3\biggl(\ln \left({(x+x_o)^2\over l x }\right)
+\Theta-\half\biggr)\biggr]
,\label{103}\eea
\be
T_{uv}=8{{\cal K}\over l^4}{(x^2-x_o^2)^2\over x^2}.
\label{104}\ee
\par
Evidently,  the approach of this section is a
substantial improvement over the prior  limiting procedure.
All properties (except arbitrary  temperature) are now  well defined 
and  all  local 
 quantities are   regular throughout the relevant  manifold.
Furthermore, the stress tensor  satisfies  the
horizon regularity conditions (\ref{CF}),  which  implies that our
choice of boundary conditions (\ref{92})  appropriately
 describes an  extremal Hartle-Hawking state.

\section{Conclusion}
In the preceding sections, we have  examined numerous
properties of a spinning BTZ black hole in  a state of  
thermal equilibrium.   An analytical description 
  of the
one-loop back reaction  was formulated  with the application of  
 perturbative techniques to a dimensionally reduced model.
The  one-loop thermodynamics were extracted   from
the on-shell Euclidean action, which effectively describes the partition
function in a semi-classical regime. When  considerations were limited to 
non-extremal  black holes,  we found
these  geometrical and thermodynamic   calculations   to be       
both  regular and unambiguously defined. 
However, 
    the extremal limit of these calculations  was shown to be 
     plagued by divergent behavior.  
This  extremal breakdown
in the one-loop approximation is suggestive of a generalized third law
of thermodynamics that prohibits      continuous evolution from non-extremal
to extremal states.
\par
As an alternative to the  limiting procedure,
we have also  considered the extremal case from  the following viewpoint:
 extremal and non-extremal black holes are qualitatively distinct
entities. In this alternative approach, the extremal solution was assumed
from the beginning and horizon regularity (in the  one-loop geometry)
was used to fix the boundary conditions. With this procedure, we found
all calculations to be regular and all  thermodynamic  properties
(with one exception) to be 
well-defined.
The one exception was   temperature, 
which we   justifiably   regarded as   an arbitrary quantity.
 Other notable results were a vanishing entropy and the horizon regularity
of the stress tensor 
in 
 the free-falling frame.
Although this analysis was limited  to the study of  BTZ black holes,
qualitatively  similar outcomes
have  been obtained for the Reissner-Nordstrom case \cite{12.99}.
 \par
The arbitrary nature of the  extremal temperature is somewhat unsettling
inasmuch as 
  the physical state is  (at least in some sense) 
a thermal one  with non-vanishing asymptotic radiation. To help clarify
this apparent  conflict, 
we take note of recent findings by Liberati et al. \cite{12}.
They have 
 considered an extremal Reissner-Nordstrom black hole undergoing collapse
and demonstrated  that (in spite of asymptotic particle production)
the temperature remains undefinable on account of a non-Planckian
distribution. Although this result does not apply directly to
static BTZ black holes, it does imply an intrinsic elusiveness in measuring the
 temperature of an extremal black hole.
\par
One may find it  intuitively disturbing  to assign
  a vanishing entropy     to 
 a  macroscopic object that  emits radiation, although a 
  strong case for this    has been   recently  put 
forth.  
 Hod \cite{11} argued in favor of  $S_{ext}=0$  by  appealing  to the
second law of thermodynamics on the basis of a {\it Gedanken} experiment. 
 However,
before any  definitive viewpoint can be reached on this subject,
 we will ultimately  require a clearer picture of what degrees of 
freedom underlie   black hole entropy  
\par
A couple of  final notes regarding our results are  in order.
Firstly,  by imposing an axially
symmetric reduction on the matter action, we have limited considerations
to the ``S-waves'' of the matter fields. Hence, from a  3-d  point of view,
the quantum effective action is only an approximation.
That is, modifications  may be required if we are to apply
these results (and conclusions) directly to 2+1-BTZ black holes 
 \cite{NO1,FZ,39}. 
However, from the viewpoint  of   1+1-dilaton  black holes, 
this  dimensional-reduction anomaly   can be considered 
as inconsequential.
\par
Finally, we again   point out  the omission of non-local terms in 
 the conformally invariant portion of the effective action
(see the discussion following Eq.(\ref{4})).  The inclusion of these
terms  would likely modify the quantitative  details
of our one-loop calculations.  
We expect, however, that the qualitative outcomes
of this paper (regarding singular extremal behavior) will   persist even
when the complete effective action is  incorporated. We hope to
formally address this issue
 in a future work. In any event, the
techniques of our  current analysis should prove useful in
 subsequent studies on  both  extremal and 
non-extremal black hole thermodynamics.

\section{Acknowledgements}
\par
 This work
was supported in part by the Natural Sciences and Engineering
Research
Council of Canada.
A.J.M.M.  would like to thank George Ellis and the U.C.T. Cosmology
group for being a gracious host  during this venture. 
 G.K. would like to thank J. Gegenberg for helpful
conversations. 
  \par

\end{document}